\documentclass[11pt]{article}

\usepackage{graphicx}
\usepackage{bm}
\usepackage{float}
\usepackage[bottom=3cm, top=3cm]{geometry}
\usepackage{hyperref}
\usepackage{siunitx}
\newcommand{\rtHz}{\sqrt{\si{Hz}}}
\newcommand{\uni}[1]{\ {\rm #1}}
\usepackage{booktabs}
\usepackage{physics}
\usepackage{cite}

\title{\vspace{-1cm}\textbf{Wavefront Sensing with a Coupled Cavity\\for Torsion-Bar Antenna}} 
\author
{Yuka Oshima$^{1}$, Satoru Takano$^{1}$, Ching Pin Ooi$^{1}$,
Yuta Michimura$^{2,3}$, Masaki Ando$^{1,3}$\\
\\
\normalsize{$^{1}$ Department of Physics, University of Tokyo, Bunkyo, Tokyo 113-0033, Japan}\\
\normalsize{$^{2}$ LIGO, California Institute of Technology, Pasadena, California 91125, USA}\\
\normalsize{$^{3}$ Research Center for the Early Universe, University of Tokyo, Bunkyo, Tokyo 113-0033, Japan}\\
\\
}

\usepackage{fancyhdr}
\begin{document} 
\date{\today} 
\baselineskip14pt  
\maketitle 
\thispagestyle{fancy}

\begin{abstract}
Torsion-Bar Antenna (TOBA) is a ground-based gravitational wave detector using torsion pendulums.
TOBA can detect intermediate-mass black hole binary mergers, gravitational wave stochastic background, and Newtonian noise,
and is useful for earthquake early warning.
A prototype detector Phase-III TOBA with 35 cm-scale pendulums
is under development to demonstrate noise reduction.
The target strain sensitivity is set to $1\times10^{-15}\uni{/\rtHz}$ between 0.1 Hz--10 Hz.
A new scheme of wavefront sensing with a coupled cavity
was proposed to measure the pendulum rotation as low as $5\times10^{-16}\uni{rad/\rtHz}$ for Phase-III TOBA.
In our method, an auxiliary cavity is used to enhance the first-order Hermite--Gaussian mode in a main cavity.
Experimental demonstration is ongoing
to confirm the feasibility of angular signal amplification and establish a method for locking a coupled cavity.
We evaluated the performance of the coupled cavity 
and concluded that angular signal amplification would be feasible with this sensor.
The coupled cavity was successfully locked to the resonance 
by the Pound--Drever--Hall technique with two modulation frequencies.
\end{abstract}

{DOI: add here the DOI provided by the GRASS LOC}

\section*{1. Introduction}

Currently operating laser interferometric gravitational wave detectors,
Advanced LIGO, Advanced Virgo, and KAGRA,
are sensitive between 10 Hz--1 kHz \cite{LIGO, Virgo, Somiya2012, Aso2013}
and successfully detected gravitational waves from solar-mass black hole and neutron star binary mergers \cite{GWTC-3}.
To observe gravitational waves in the lower frequency range (0.1 Hz--10 Hz), 
various kinds of detectors have been proposed and are being developed.
These detectors are expected to observe
intermediate-mass black hole binary mergers and the gravitational wave stochastic background. 
One possible way to obtain highly sensitive detectors between 0.1 Hz--10 Hz is to use spacecrafts
such as LISA and DECIGO \cite{LISA, DECIGO}.
Spacecrafts are free from seismic noise and can be regarded as free masses in all frequencies.
However, the cost of development is expensive and the maintenance is difficult.
Another way is to develop ground-based detectors using a different principle from currently operating detectors.
TOBA is composed of torsion pendulums and 
we aim to detect the rotational motion of said torsion pendulums caused by gravitational waves \cite{Ando2010}.
The target strain sensitivity of TOBA is set to $1\times10^{-19}\uni{/\rtHz}$ between 0.1 Hz and 10 Hz.

We are currently developing the third prototype detector named Phase-III TOBA,
which has the sensitivity of $1\times10^{-15}\uni{/\rtHz}$  \cite{Shimoda2020}.
A highly sensitive angular sensor is needed
to read out the angle of torsion pendulums from as low as $5\times10^{-16}\uni{rad/\rtHz}$.
Michelson interferometers are often used as angular sensors by putting two-arm mirrors on the edge of the bars,
but they do not have sufficient sensitivity \cite{Ishidoshiro2011, Shoda2017, Ross2021}.
Alternatively, wavefront sensors are sensitive angular sensors with optical cavities \cite{Morrison1994a, Morrison1994b}.
However, conventional wavefront sensors also cannot achieve the requirement
even with high laser power and large beam spots \cite{Aso2013}.
Therefore, we proposed a new type of angular sensor, a wavefront sensor with a coupled cavity \cite{Shimoda2022}.
In our method, we build an auxiliary cavity behind the main cavity to enhance the angular signal.

In this paper,
we review both the proposed and prototype detectors of TOBA in Section 2.
Section 3 explains the principle and simulation results of a wavefront sensor with a coupled cavity.
Section 4 reports the status of the experimental demonstration 
and Section 5 concludes the paper.


\section*{2. Torsion-bar antenna}

\begin{figure} [H]
\centering
\includegraphics[width=10cm]{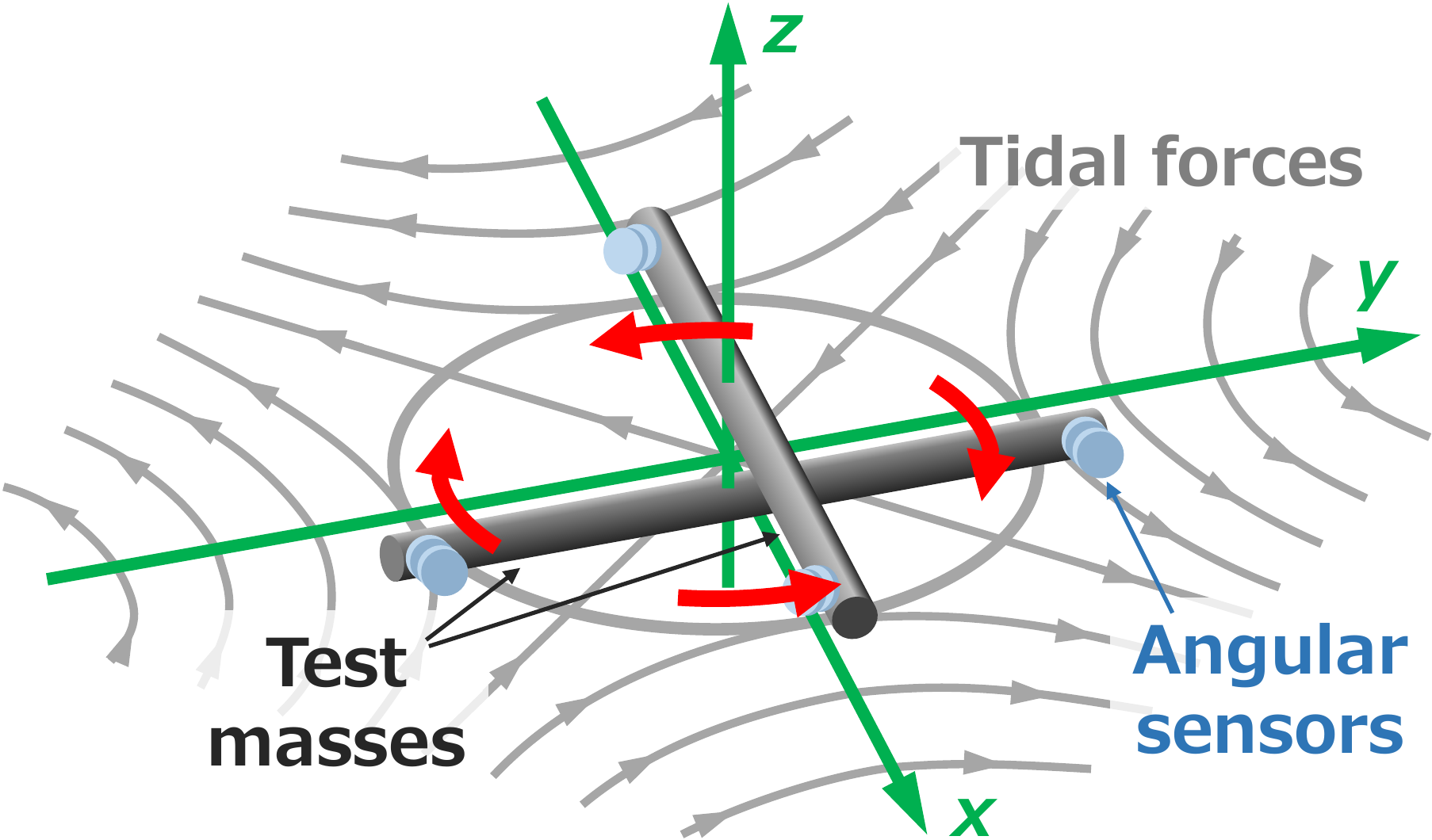}
\caption{\small{The configuration of TOBA.
Two test mass bars are suspended on the $x-y$ plane on the ground.
Gray arrows show tidal forces by gravitational waves
and red arrows represent the differential motion of test masses by tidal forces.
Angular sensors with optical cavities are attached to the edges of bars.
}}\label{fig:TOBA}
\end{figure}

TOBA was proposed to detect gravitational waves between 0.1 Hz--10 Hz \cite{Ando2010}.
The target sensitivity of TOBA is set to $1\times10^{-19}\uni{/\rtHz}$.
The configuration of TOBA is shown in Figure \ref{fig:TOBA}.
TOBA is composed of two test mass bars suspended horizontally on the ground.
We aim to detect the torsional rotation of TOBA caused by tidal forces due to gravitational waves.
TOBA has better sensitivity in low frequency compared to current detectors
since the resonant frequency of torsional motion is lower than that of translational motion.

The science of TOBA can be divided into two aspects; astrophysics and geophysics.
For astrophysics,
we can observe intermediate-mass black hole binary mergers within $\sim$10 Gpc and 
gravitational wave stochastic background up to $\Omega_{\rm{GW}} \sim 10^{-7}$ \cite{Ando2010}.
TOBA will also provide us with geophysical information thanks to the ground-based configuration.
Newtonian noise is the fluctuations of the gravitational field
caused by any moving masses \cite{Saulson1984}.
Representative examples are seismic Newtonian noise \cite{Hughes1998}
and atmospheric Newtonian noise \cite{Creighton2008, Fiorucci2018}.
Newtonian noise is a candidate for the dominant noise
for third-generation detectors, Einstein Telescope, and Cosmic Explorer \cite{ET, CE}.
First direct detection of Newtonian noise is expected with TOBA \cite{Harms2013}.
In addition, we can improve earthquake early warning systems
because gravity perturbation caused by earthquakes propagates TOBA with the speed of light,
which is much faster than seismic waves, used for current warning systems \cite{Harms2015, Juhel2018}.

In order to confirm the detection principle,
we developed two prototype detectors named Phase-I TOBA and Phase-II TOBA.
We achieved the sensitivity of $1\times10^{-8}\uni{/\rtHz}$ at 0.1 Hz and 
$1\times10^{-10}\uni{/\rtHz}$ at 7 Hz, respectively,
and placed the upper limits to gravitational wave stochastic background
\cite{Ishidoshiro2011, Shoda2017, Shoda2014, Kuwahara2016}.
Now, we are developing the third prototype detector, Phase-III TOBA,
to demonstrate noise reduction \cite{Shimoda2020}.
The target sensitivity of Phase-III TOBA is set to $1\times10^{-15}\uni{/\rtHz}$ between 0.1 Hz and 10 Hz
with 35 cm bars and operation at cryogenic temperature.
After establishing Phase-III TOBA,
we plan to build the final version of TOBA
with 10 m bars at 4 K
to detect gravitational waves of $1\times10^{-19}\uni{/\rtHz}$ in the range of 0.1 Hz--10 Hz.

\section*{3. Wavefront sensor with a coupled cavity}

\subsection*{Principle of angular signal amplification}

\begin{figure} [H]
\centering
\includegraphics[width=14cm]{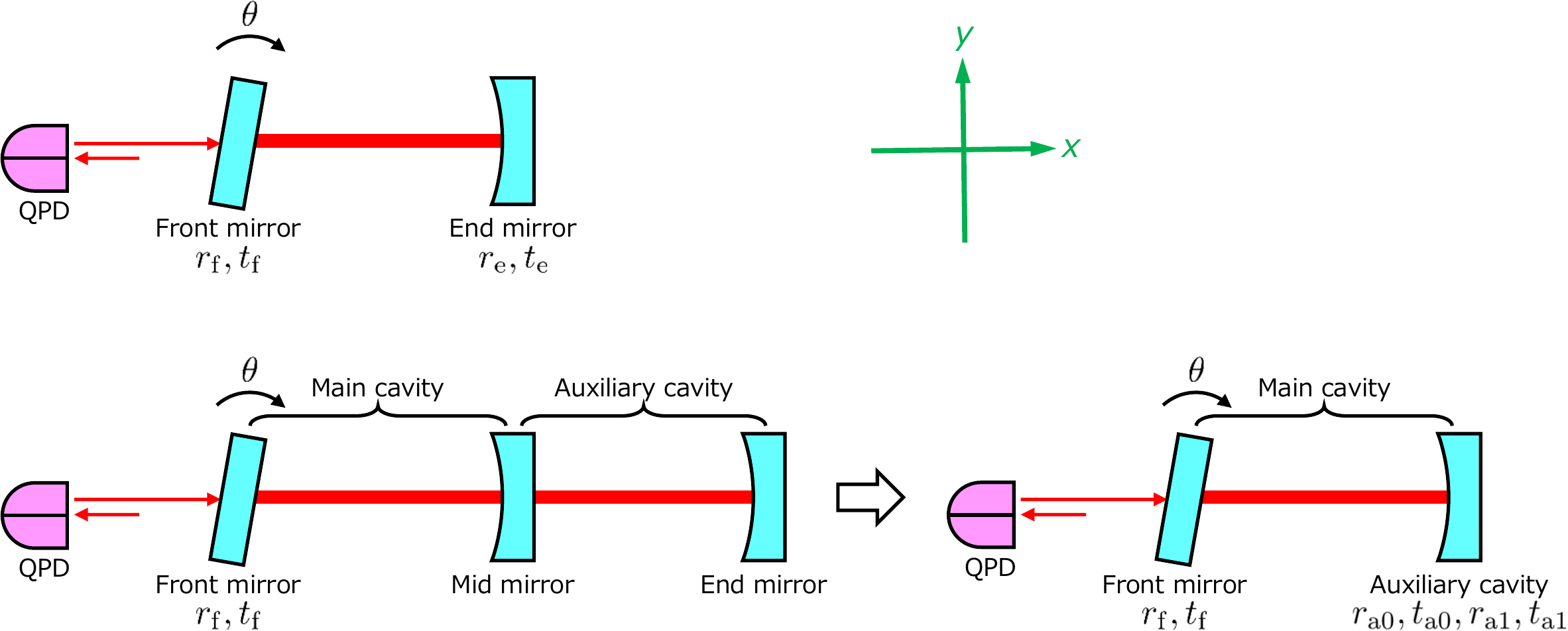}
\caption{\small{The configuration of a wavefront sensor (top),
and a wavefront sensor with a coupled cavity (bottom).
Front mirrors are misaligned by $\theta$ on the $x-y$ plane as test masses.
QPD: quadrant photodetector.
}}\label{fig:CoupledWFS}
\end{figure}

Figure \ref{fig:CoupledWFS} (top) shows the configuration of a wavefront sensor.
A wavefront sensor consists of an optical cavity, that is two facing high reflective mirrors.
Ideally, only the fundamental Hermite--Gaussian mode (HG$_{00}$ mode) exists in the cavity.
However, when the mirror of the cavity is misaligned on the $x-y$ plane,
the first-order HG mode (HG$_{10}$ mode) is created from part of the HG$_{00}$.
By detecting the interference between HG$_{00}$ and HG$_{10}$ modes
at a reflection port with a quadrant photodetector
and taking the difference between left and right signals,
we can measure the angle of the mirror.
When the front mirror of the cavity is misaligned,
the complex reflectivity of the cavity for HG$_{10}$ mode is calculated by
\begin{equation}
r_{\rm c1} = \beta r_{\rm f} + 
\frac{\beta t_{\rm f}^2 r_{\rm f} r_{\rm e}^2 e^{-i\phi_0}e^{-i\phi_1}}{\qty(1 - r_{\rm f}r_{\rm e}e^{-i\phi_0}) \qty(1 - r_{\rm f}r_{\rm e}e^{-i\phi_1})},
\end{equation}
where $r_{\rm f}, r_{\rm e}, t_{\rm f}$, and $t_{\rm e}$
are the reflectivity and transmissivity of the mirrors
with ${\rm f}$ and ${\rm e}$ denoting the front and end mirrors, respectively, 
and $\phi_0$ and $\phi_1$ are the intracavity phase for HG$_{00}$ and HG$_{10}$ modes, respectively.
$\beta$ is defined by $2i\theta / \alpha_0$, where $\theta$ is the tilt angle of the front mirror
and $\alpha_0$ the divergence of the laser beam.
When HG$_{00}$ mode is resonant in the cavity,
$\phi_0=0$ and $\phi_1= - \zeta$ are satisfied, 
where $\zeta$ is the round-trip Gouy phase.
$r_{\rm c1}$ is written as 
\begin{equation}
r_{\rm c1} = \beta r_{\rm f} + 
\frac{\beta t_{\rm f}^2 r_{\rm f} r_{\rm e}^2 e^{i\zeta}}{\qty(1 - r_{\rm f}r_{\rm e}) \qty(1 - r_{\rm f}r_{\rm e}e^{i\zeta})}.
\end{equation}
The amplification of HG$_{00}$ mode can be seen as the term of $1/(1 - r_{\rm f}r_{\rm e})$.
In general, HG$_{10}$ mode is not amplified in the cavity
since the round-trip Gouy phase meets
$1/(1 - r_{\rm f}r_{\rm e}e^{i\zeta_{\rm round}})\simeq 1$
due to $0<\zeta<2\pi$.

\begin{figure} [H]
\centering
\includegraphics[width=8cm]{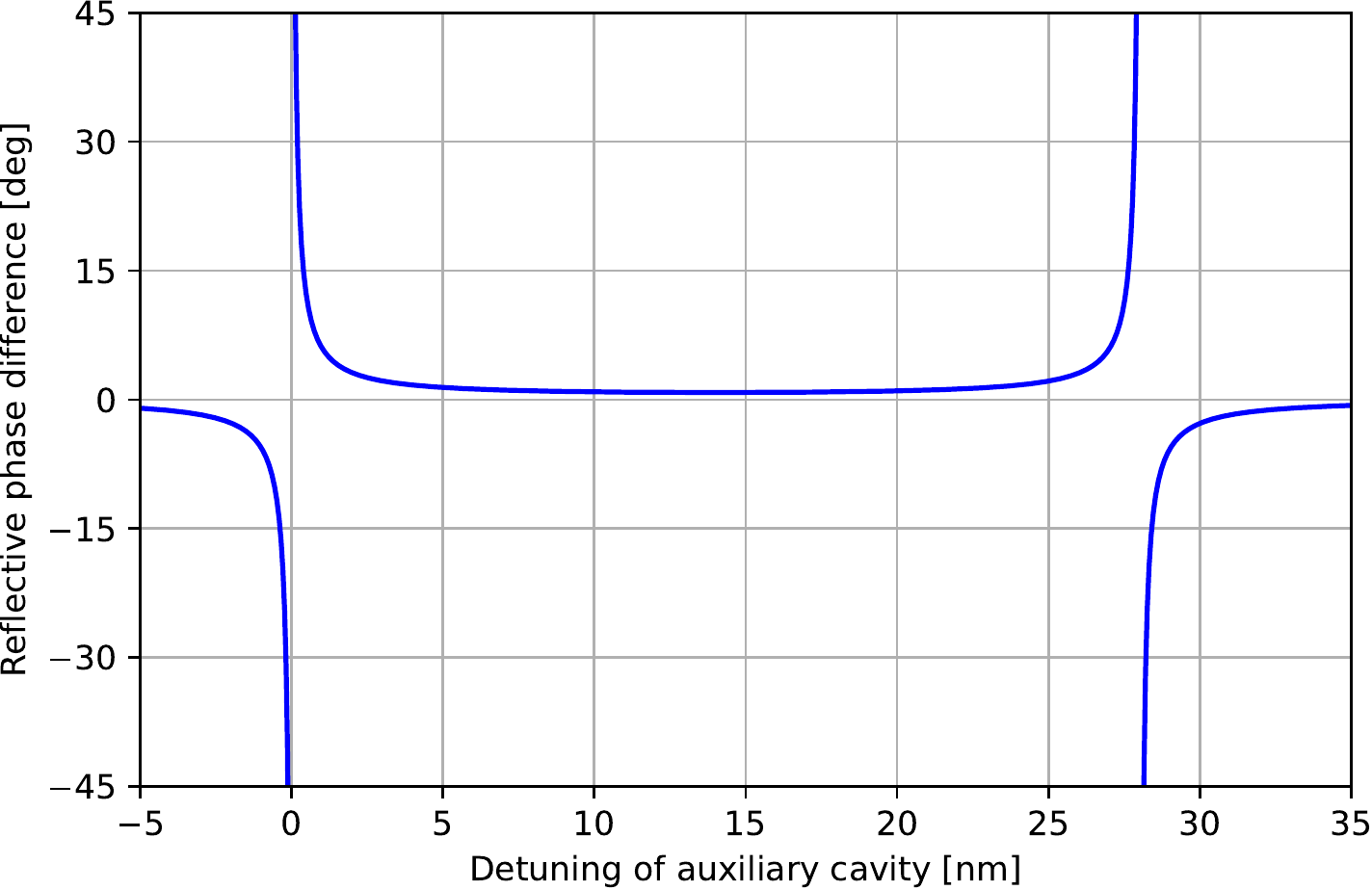}
\caption{\small{One example of the reflective phase difference of an auxiliary cavity
between HG$_{00}$ and HG$_{10}$ modes.
}}\label{fig:PhaseCompensation}
\end{figure}

Figure \ref{fig:CoupledWFS} (bottom) shows
the configuration of a new type of angular sensor, a wavefront sensor with a coupled cavity.
We put an additional mirror behind the main cavity to create an auxiliary cavity.
The auxiliary cavity can be regarded as a compound mirror with complex reflectivity and transmissivity.
When the front mirror is misaligned,
the complex reflectivity at the front mirror for HG$_{10}$ mode is given by
\begin{equation}
r_{\rm cc1} 
= \beta r_{\rm f} + 
\frac{\beta t_{\rm f}^2 r_{\rm f} r_{\rm a0} r_{\rm a1} e^{-i\phi_{\rm m0}}e^{-i\phi_{\rm m1}}}{\qty(1 - r_{\rm f}r_{\rm a0}e^{-i\phi_{\rm m0}}) \qty(1 - r_{\rm f}r_{\rm a1}e^{-i\phi_{\rm m1}})},
\end{equation}
where $r_{\rm a0}, r_{\rm a1}, t_{\rm a0}$, and $t_{\rm a1}$
are the reflectivity and transmissivity of the auxiliary cavity for HG$_{00}$ and HG$_{10}$ modes, respectively, and 
$\phi_{\rm m0}$ and $\phi_{\rm m1}$ are the intracavity phase
of the main cavity for HG$_{00}$ and HG$_{10}$ modes, respectively.
The auxiliary cavity gives a different reflective phase shift
for HG$_{00}$ and HG$_{10}$ modes in the main cavity
as shown in Figure \ref{fig:PhaseCompensation}.
Therefore, the Gouy phase differences between $\phi_{\rm m0}$ and $\phi_{\rm m1}$ can be canceled 
by tuning the length of the auxiliary cavity
and HG$_{00}$ and HG$_{10}$ modes can be resonant at the same time in the main cavity.
When both HG$_{00}$ and HG$_{10}$ modes are resonant in the main cavity,
$r_{\rm cc1}$ is described by 
\begin{equation}
r_{\rm main1} 
= \beta r_{\rm f} + 
\frac{\beta t_{\rm f}^2 r_{\rm f} |r_{\rm a0}| |r_{\rm a1}| }{\qty(1 - r_{\rm f} |r_{\rm a0}|) \qty(1 - r_{\rm f}|r_{\rm a1}|)}.
\end{equation}
Not only HG$_{00}$ mode but also HG$_{10}$ mode is enhanced 
if $|r_{\rm aux0}| \sim 1$ and $|r_{\rm aux1}| \sim 1$ are satisfied.
As a result, the angular signal intensity,
which is created from the interference between these two modes,
with a coupled cavity can be larger than that with a single cavity.

\subsection*{Simulation with FINESSE}

We worked on the numerical calculation 
to confirm the angular signal amplification with a coupled cavity.
The simulation software FINESSE was used in this calculation
since the configuration of a coupled cavity is complicated.
We confirmed that FINESSE worked well
by comparing the simulation result with the analytical solution of a single cavity.

\begin{figure} [H]
\begin{minipage}{0.5\linewidth}
\centering
\includegraphics[width=75mm]{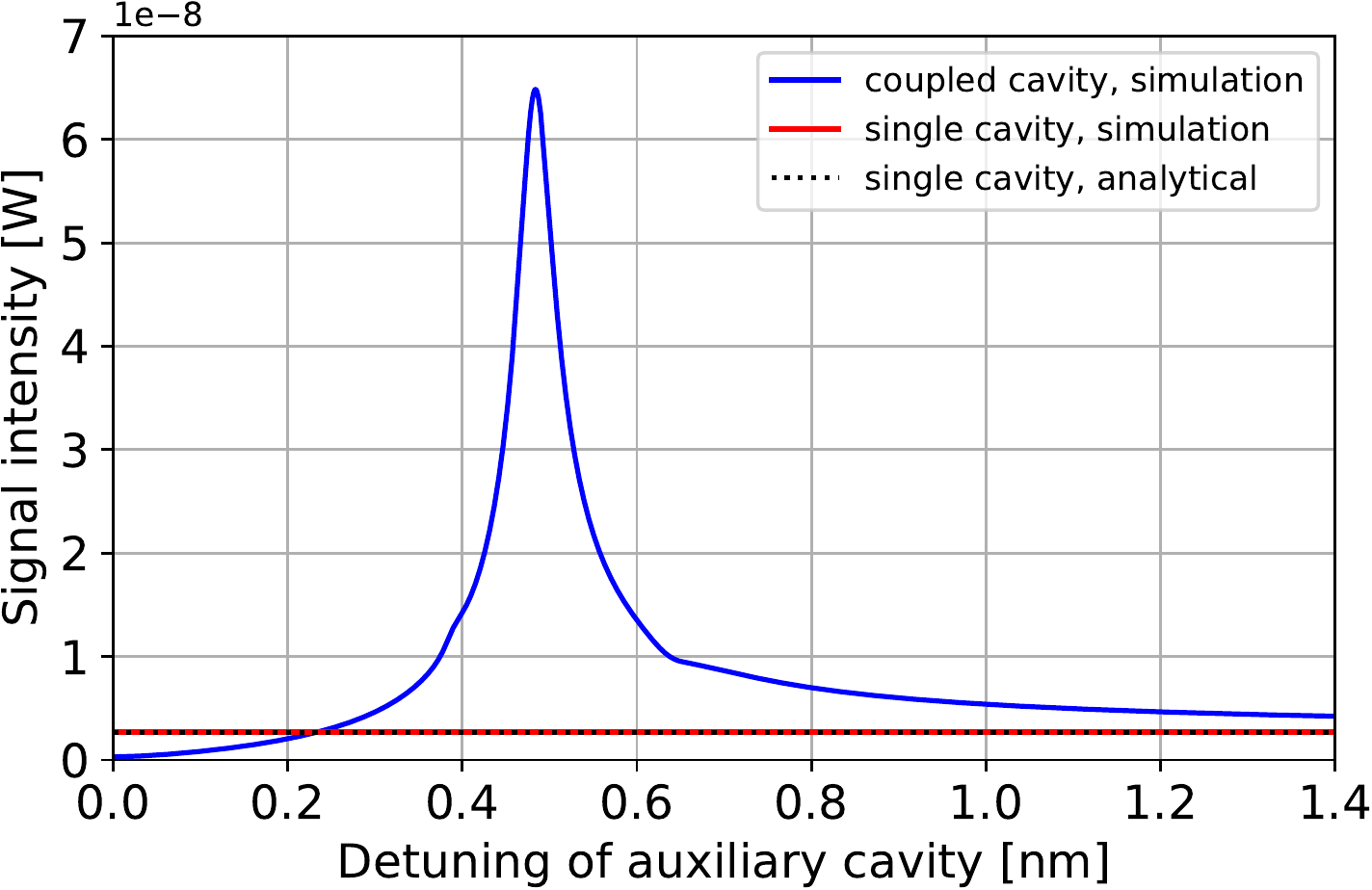}
\end{minipage}
\hfill
\begin{minipage}{0.5\linewidth}
\centering
\includegraphics[width=75mm]{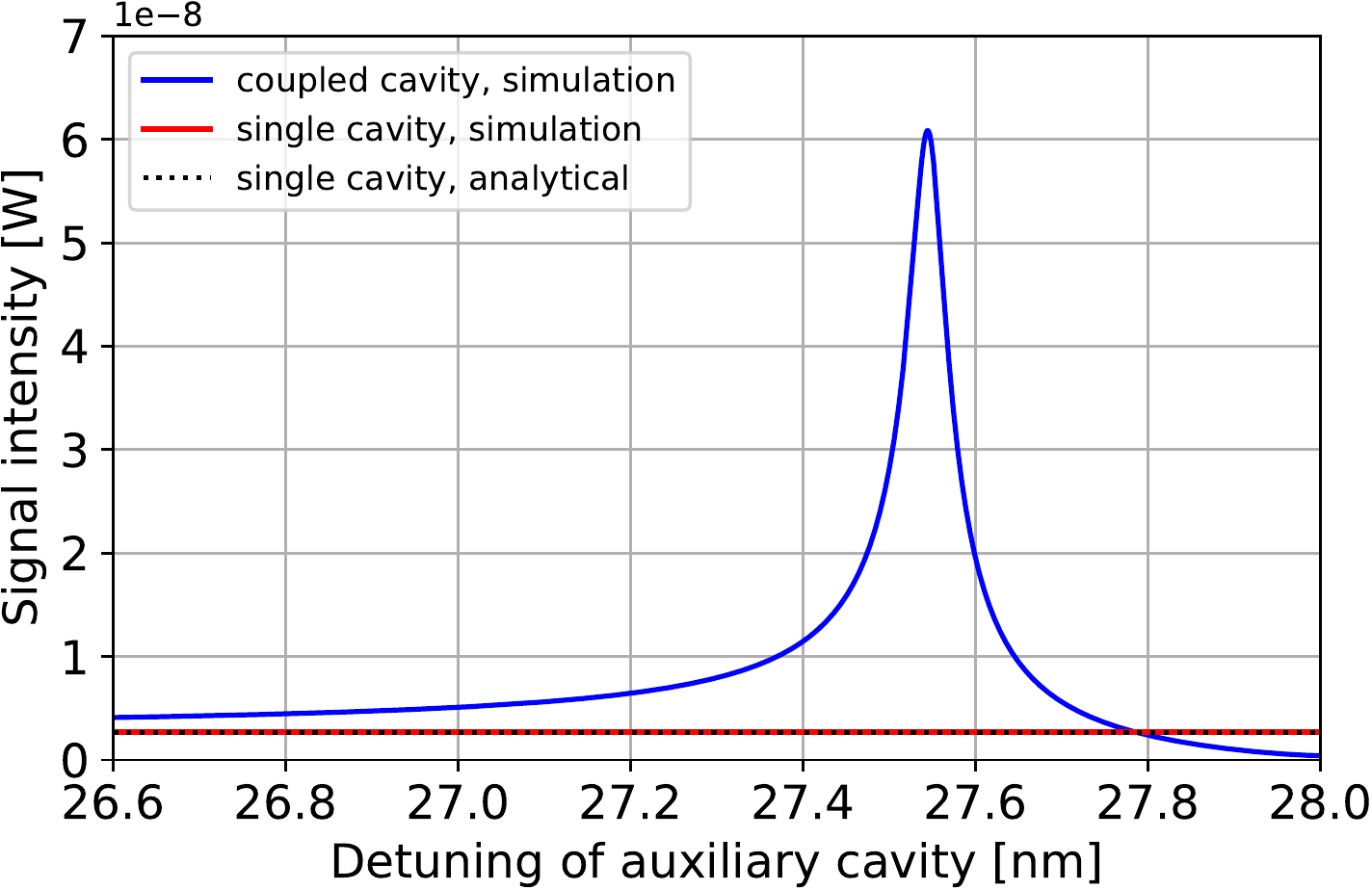}
\end{minipage}
\caption{\small{The simulation results for angular signal amplification with a coupled cavity 
near HG$_{00}$ mode resonance (left), and near HG$_{10}$ mode resonance (right) in the auxiliary cavity.
The horizontal axis shows the detuning length of the auxiliary cavity from HG$_{00}$ mode resonance
and the vertical axis shows the angular signal intensity.
The black dotted line is the analytical solution with a single cavity.
The blue (red) solid lines represent
the simulation result with a coupled cavity (a single cavity).
}}\label{fig:FINESSE}
\end{figure}

A coupled cavity with a length of 15 cm and finesse of 350 was simulated.
The front mirror was misaligned by 1 nrad and the power of the laser beam was 1 W.
Two quadrant photodetectors were placed 
at the reflection port with the Gouy phase separation of $\pi/2$.
The signal intensity was calculated for the square sum of two photodetector outputs
sweeping the auxiliary cavity length.
The simulation results are shown in Figure \ref{fig:FINESSE}. 
Simultaneous resonance of HG$_{00}$ and HG$_{10}$ modes in the main cavity happens
at two points; near HG$_{00}$ and HG$_{10}$ modes resonance in the auxiliary cavity.
At those points, angular signal amplification could be seen compared to conventional wavefront sensors.  

\section*{4. Experimental demonstration}

\subsection*{Experimental setup}

\begin{figure} [H]
\begin{minipage}{0.5\linewidth}
\centering
\includegraphics[width=90mm]{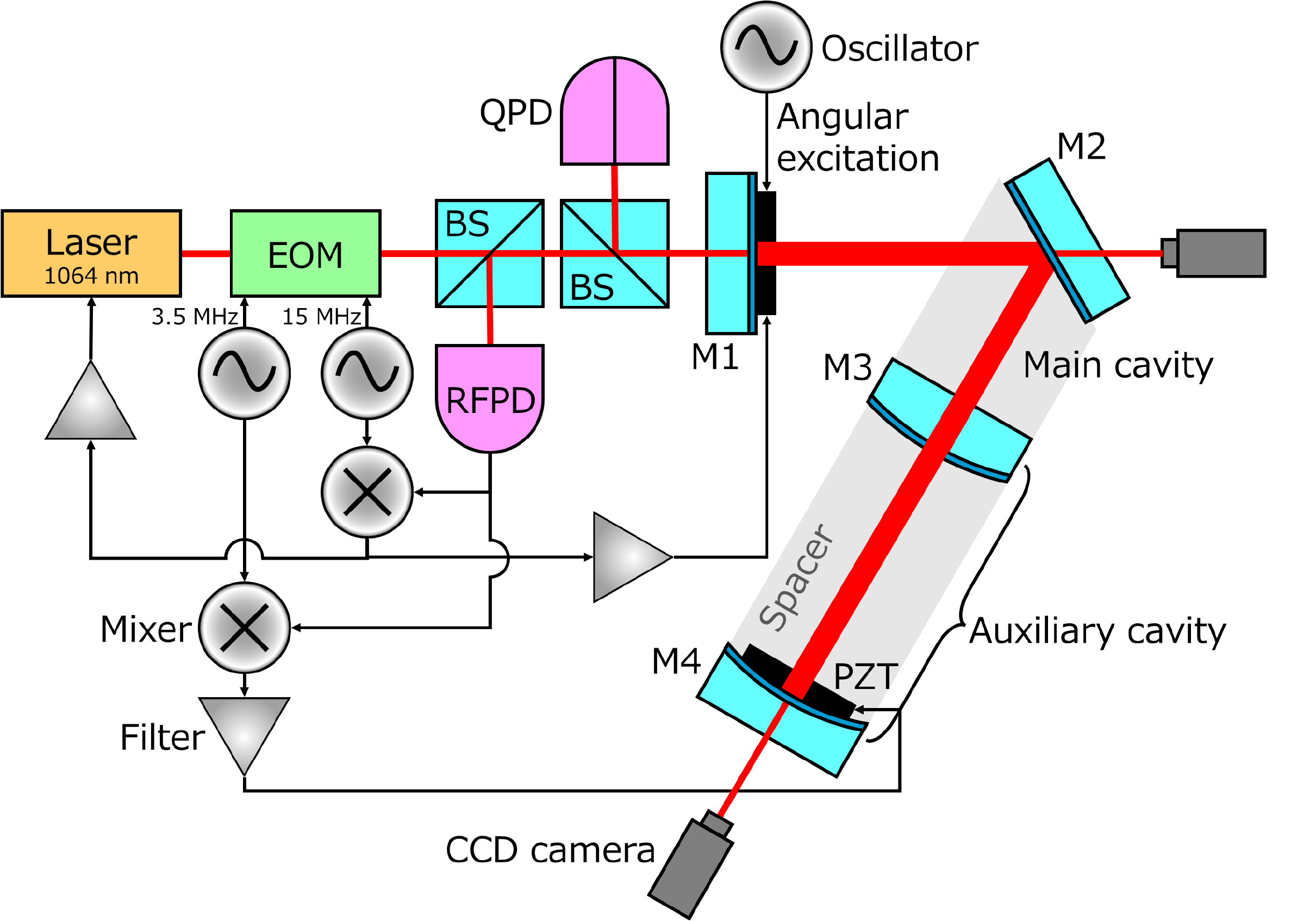}
\end{minipage}
\hfill
\begin{minipage}{0.5\linewidth}
\centering
\includegraphics[width=50mm]{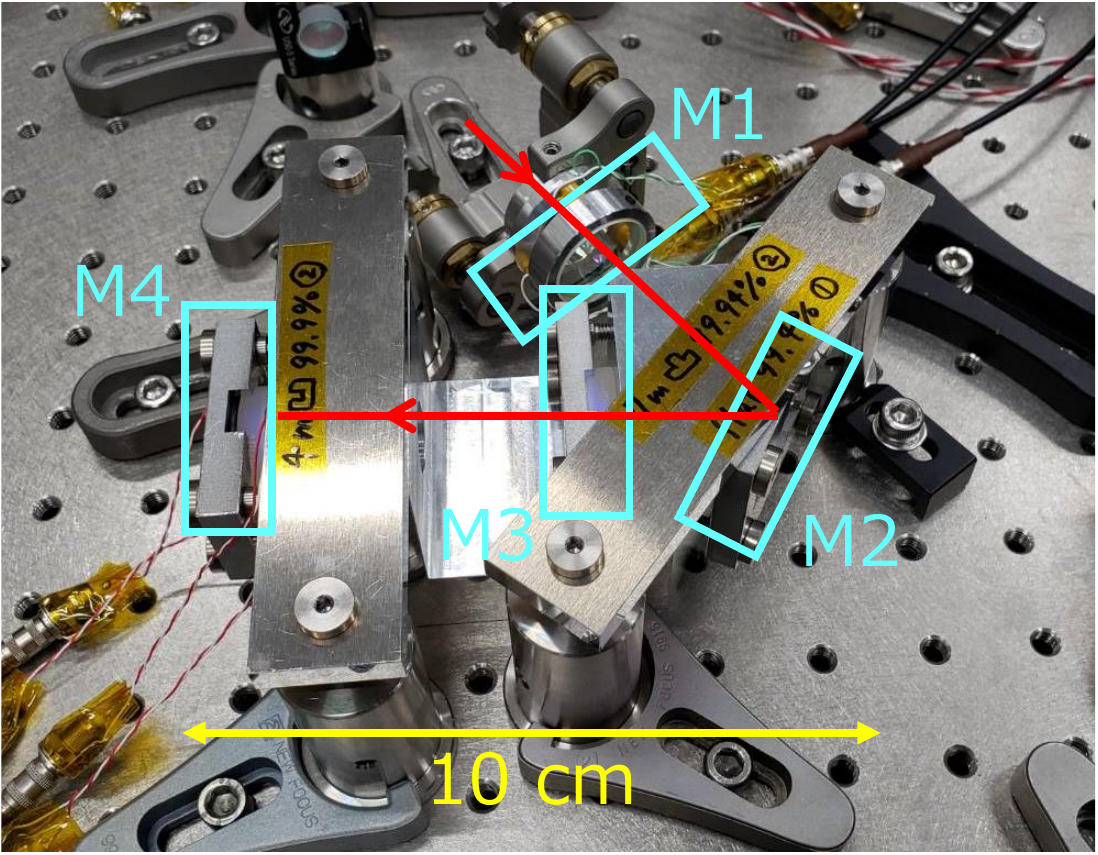}
\end{minipage}
\caption{\small{The schematic of the experimental demonstration 
for a wavefront sensor with a coupled cavity (left), and a picture of the coupled cavity (right).
The dark blue region of the mirrors M1--M4 indicates high reflection coatings.
EOM: electro-optic modulator. BS: beam splitter.
QPD: quadrant photodetector.
RFPD: radio frequency photodetector.
PZT: piezoelectric actuators.
}}\label{fig:setup}
\end{figure}

Experimental demonstration for a wavefront sensor with a coupled cavity
is underway to evaluate signal amplification and establish a method for locking a coupled cavity.
Figure \ref{fig:setup} (left) shows the simplified schematic of the experiment.
The laser beam with the wavelength of 1064 nm and the power of 3 mW was fed into the coupled cavity.
The main cavity was folded with one more mirror to monitor the transmitted light directly with a CCD camera.
The incident angle at the mirror M2 was 30 deg.
The mirrors M2--M4 were rigidly fixed to a spacer made of aluminum to stabilize the alignment.
The front mirror M1 was used as a test mass by injecting angular excitation signals.
A photo of the coupled cavity is shown in Figure \ref{fig:setup} (right).

\subsection*{Design and evaluation of the coupled cavity parameters}

Designed parameters of the coupled cavity are shown in Table \ref{tab:parameters}.
The coupled cavity was designed so that Gouy phase compensation was enabled and 
eigenmodes of the two cavities matched.
Minimizing the loss of the auxiliary cavity  
is critical for Gouy phase compensation.
Therefore, high reflection coatings were facing the auxiliary cavity
and we designed a coupled cavity to compensate for phase difference
even if a loss of up to 0.1\% exists in the auxiliary cavity.
All four mirrors M1--M4 were custom-made by Layertec.
Uncertainties in power reflectivity and radius of curvature were provided by Layertec,
and listed under designed values in Table \ref{tab:parameters}.

We evaluated the performance of the cavities
by modulating the laser frequency and taking the cavity scan of transmitted light.
Results are summarized in Table \ref{tab:parameters}.
The measured finesse of the main cavity 
was smaller than the designed finesse.
This might be caused by the optical loss from the mirror M3
since the laser beam was injected from the opposite side of the high reflection coating
and passed through the anti-reflection coating and substrate.
Note that the smaller finesse of the main cavity 
is independent of whether the auxiliary cavity can compensate for the Gouy phase difference,
whereas it is dependent on the angular signal amplification ratio. 
The measured finesse of the auxiliary cavity 
and Gouy phase for the two cavities were consistent with the designed ones.
We concluded that the auxiliary cavity would be able to compensate for phase difference
and enhance the angular signal in the main cavity. 

\begin{table} [H]
\caption{\label{tab:parameters}Summary of the parameters of the coupled cavity.}
\begin{center}
\begin{tabular}{lll}
\toprule
&Designed values&Measured values\\
\midrule
Power reflectivity of M1 and M2 & $99.4\pm0.3\%$ & -- \\
Power reflectivity of M3 & $99.94\pm0.02\%$ & -- \\
Power reflectivity of M4 & $99.90\pm0.02\%$ & -- \\
Radius of curvature of M1 and M2 & $\infty$ (flat) & -- \\
Radius of curvature of M3 & $\SI{7}{m} \pm 2\%$ (convex) & -- \\
Radius of curvature of M4 & $\SI{4}{m} \pm 1\%$ (concave) & -- \\
Length of the main cavity & \SI{78.7}{mm} & -- \\
Length of the auxiliary cavity & \SI{62.5}{mm} & -- \\
Finesse of the main cavity & $337^{+330}_{-112}$ & $(2.0 \pm 0.2) \times 10^2$ \\
Finesse of the auxiliary cavity & $(3.93^{+1.30}_{-0.79}) \times 10^3$ & $(4.1 \pm 0.2) \times 10^3$  \\
Round-trip Gouy phase of the main cavity & $12.2 \pm 0.1 \uni{deg}$  & $12.1 \pm 1.0$~deg \\
Round-trip Gouy phase of the auxiliary cavity & $9.49^{+0.22}_{-0.24} \uni{deg}$ & $9.54 \pm 0.04$~deg \\
Mode matching ratio of the main cavity & -- & $87 \pm 2\%$ \\
Mode matching ratio of the auxiliary cavity & -- & $94 \pm 2\%$ \\
\bottomrule
\end{tabular}
\end{center}
\end{table}

\subsection*{Cavity locking scheme}

Figure \ref{fig:setup} (left) shows the method for locking the coupled cavity. 
The coupled cavity was locked to the resonance by the Pound--Drever--Hall (PDH) technique \cite{Drever1983, Black2001}.
A different method was demonstrated in \cite{Shimoda2022};
the main cavity was locked with the PDH technique
and the auxiliary cavity length was locked by keeping the transmitted power constant.
This method was easier to operate but failed to lock the auxiliary cavity to the resonance.

Utilizing the PDH technique,
the two modulation frequencies chosen were
15 MHz for the main cavity and 3.5 MHz for the auxiliary cavity.
Hierarchical control was introduced for the main cavity to prevent transmitting disturbances
from the main cavity to the auxiliary cavity through the laser frequency;
the feedback signal for the main cavity below 20 Hz
was fed back to the piezoelectric actuator in the laser source,
while that under 20 Hz was fed back to the piezoelectric actuator on the mirror M1.
The feedback signal for the auxiliary cavity was fed back to the piezoelectric actuator on the mirror M4.
The main and auxiliary cavities were successfully locked simultaneously
by the PDH technique with the two modulation frequencies.


\section*{5. Conclusions}

Torsion-Bar Antenna (TOBA) is a ground-based gravitational wave detector
using 10 m torsion pendulums.
The resonant frequency of torsional motion is $\sim1\uni{mHz}$,
allowing TOBA to have a design sensitivity of $1\times10^{-19}\uni{/\rtHz}$ between 0.1 Hz and 10 Hz. 
TOBA can detect intermediate-mass black hole binary mergers, gravitational stochastic background, Newtonian noise, and earthquakes.
A prototype detector Phase-III TOBA with 35 cm pendulums is under development 
to demonstrate noise reduction.
The target sensitivity is set to $1\times10^{-15}\uni{/\rtHz}$ in the range of 0.1 Hz--10 Hz.

To achieve our target sensitivity, we need to measure the pendulum rotation precisely.
The requirement of shot noise for Phase-III TOBA is $5\times10^{-16}\uni{rad/\rtHz}$.
A new angular sensor, a wavefront sensor with a coupled cavity, was proposed for Phase-III TOBA.
An auxiliary cavity is used to compensate for the Gouy phase difference of the main cavity 
and enhance the HG$_{10}$ mode in the main cavity.

Experimental demonstration of a wavefront sensor with a coupled cavity is ongoing 
to evaluate signal amplification and establish a method for locking a coupled cavity.
The assembly of the optics as well as the performance evaluation of the coupled cavity has been completed.
Measured parameters critical to phase compensation were consistent with the designed ones.
The coupled cavity was locked by the PDH technique with the two modulation frequencies.
We are working on further demonstration to evaluate signal amplification quantitatively.

\section*{Acknowledgments}

We would like to thank Shigemi Otsuka and Togo Shimozawa for manufacturing the mechanical parts used. 
This work was supported by
MEXT Quantum LEAP Flagship Program
(MEXT Q-LEAP) Grant Number JPMXS0118070351.
Y.O. was supported by JSPS KAKENHI Grant Number JP22J21087 and JSR Fellowship, the University of Tokyo.

\end{document}